\documentclass[aps,prd,10pt,a4paper,altaffilletter,amssymb,showpacs,nofootinbib,twocolumn,showkeys,floatfix]{revtex4-1}

\usepackage{color}
\usepackage[mathscr]{euscript}
\usepackage{graphicx}
\usepackage{amsmath}
\usepackage{array}
\usepackage{booktabs}
\usepackage{multirow}

\renewcommand{\arraystretch}{1.4}

\begin{document}

\title{A possible geometrical origin of the accelerated expansion of the universe}
\author{Ariadna Montiel$^{1,2}$,  Nora Bret\'on$^{2}$ , Rub\'en Cordero$^{3}$ and Efra\'{\i}n Rojas$^{4}$}
\affiliation{$^{1}$Depto. de F\'{\i}sica y  Matem\'aticas, Tecnol\'ogico de Monterrey, Campus Estado de M\'exico, Atizap\'an de Zaragoza, Estado de M\'exico,  Apdo. 52926, M\'exico \\
$^{2}$Depto. de F\'{\i}sica, Centro de Investigaci\'on y de Estudios Avanzados
del I. P. N., Apdo. 14-740, D.F., M\'exico.\\
$^{3}$Departamento de F\'{\i}sica, Escuela Superior de F\'{\i}sica y Matem\'aticas
del IPN, Unidad Adolfo L\'opez Mateos, Edificio 9, 07738 M\'exico, Distrito Federal, M\'exico \\
$^{4}$Facultad de F\'{\i}sica, Universidad Veracruzana, 9100 Xalapa, Veracruz, M\'exico}

\begin{abstract}
The modified geodetic brane cosmology (MGBC) is tested with observational data.
The MGBC is derived from the geodetic brane gravity action corrected by
the extrinsic curvature of the braneworld. The density parameter coming from
this additional term produces an accelerated expansion of geometrical origin.
Subject to the Supernovae Ia, Observable Hubble parameter, Baryon Acoustic Oscillations and
Cosmic Microwave Background probes, the obtained
fit provides enough evidence in the sense that the extrinsic curvature effect is able to
reproduce the accelerated expansion of the universe without need of invoking dark
energy, exotic matter or cosmological constant. Moreover the MGBC is free of the
problems present in other braneworld models.
\end{abstract}


\maketitle

\section{Introduction}

One of the current challenging problems in physics is the explanation of the
accelerated expansion of the Universe. The main hypothesis to explain this fact
relies on the existence of an unusual energy form, the so-called dark energy,
that for example can be produced by scalar fields incarnated in a great variety of field theories like
quintessence, k-essence, Galileons, etc~\cite{Amendo}. Aside, a plenty of
modified theories of gravity have been proposed to elucidate the origin of the
cosmic acceleration like $f(R)$ gravity, scalar-tensor theories, braneworlds
models, etc. Regarding this last, in the past decade an intense research on
the braneworld models has captivated cosmologists~\cite{ArkaniHamed:1998rs,Antoniadis:1998ig,Randall:1999ee,Randall:1999vf,Dvali:2000hr}.
Some of these models do have an accelerated epoch where exotic matter
is not needed.

An attempt tied only to geometry to model our accelerated Universe was suggested recently and
named modified geodetic brane cosmology~\cite{Cordero2012} (MGBC), based in the RT model~ \cite{Regge, Davidson:1999ru,Karasik:2002du}.
This turns out to be an alternative to the cosmological constant problem for describing
the late accelerated expansion of the Universe.
It can also be thought as an implementation of a gravity theory from the perspective
of string theory. In this framework gravity, through a Ricci scalar, is considered to
be induced on a hypersurface which is evolving in a higher dimensional Minkowski space
besides of taking into account a linear correction term in the extrinsic curvature of the brane trajectory.
Physically, this correction term can be considered as a remanent of the Dvali, Gabadadze and
Porrati (DGP) \cite{Dvali:2000hr} model that modifies General Relativity (GR) and exhibiting
a self-accelerated phase that may explain the current cosmological expansion.
In this sense, MGBC mimics the accelerating expansion of this type of universe.

There is a main difference with general relativity. Unlike usual gravity theories
where the metric components are considered to be fundamental variables, in MGBC the
embedding functions are the field variables instead of the four-dimensional metric components.
Within a minisuperspace framework these local coordinates induced on the hypersurface are
considered to be dynamical and form a minisuperspace in the higher dimensional embedding space.
This type of model is often called \textit{geodetic
brane gravity} (GBG) \cite{Davidson:1999ru} since it is not considered the gravitational effect of the
brane on the background spacetime.
In fact, since the point of view of the so called unified brane cosmology (UBC) discussed by Davidson and Gurwich \cite{Gurwich:2009rj}, the GBG is one part of several brane models like Randall-Sundrum (RS) \cite{Randall:1999ee,Randall:1999vf}, DGP and RT models that can be described in a unified way. In the unified brane cosmology, by means of a Dirac brane style variation, the effect of gravity produced by the brane in the bulk is considered and the Israel Junction conditions are relaxed. If the gravitational effect of the brane on the bulk is not considered the equations of motion can be obtained by taking the background gravitational constant very large. A fundamental quantity that enters into the game is the conserved bulk energy and plays the
role of a parameter that deviates the model from GR.

On geometrical grounds, the MGBC is conformed only by the first three Lovelock brane invariants
associated to the brane trajectory (our Universe). Such second-order field theories
are conformed by actions whose Lagrangian is the sum of geometric quantities constructed from the
extrinsic curvature and the induced metric 
\begin{align}
 L&= a_0+a_1K+a_2R+a_3 (K^3-3KK_{\mu \nu}K^{\mu \nu}+2K^{\mu}_{ \alpha}K^{\alpha}_{\beta}K^{\beta}_{\mu}) \nonumber \\
& +a_4(R^2-4R_{\mu \nu}R^{\mu \nu}+ R_{\alpha \beta \gamma \delta}R^{\alpha \beta \gamma \delta})+a_5{\cal O}(K^5),
 \label{Lovelock}
\end{align}
where $R$, $R_{\mu \nu}$ and $R_{\mu \nu \alpha \beta}$ 
are the Ricci scalar, Ricci tensor and the Riemann tensor constructed by the associated Gauss-Codazzi
integrability condition for surfaces, $R_{\mu \nu \alpha \beta} = K_{\mu \alpha}K_{\nu \beta}
- K_{\mu \beta} K_{\nu \alpha}$, and $K_{\mu \nu}$ is extrinsic curvature of the worldvolume,
respectively~\cite{CQG-2013}. The ensuing equations of motion from~\eqref{Lovelock} remain second-order
which leads to a ghost-free field theory.

In spite of its theoretical predictions, MGBC has not been observationally tested. This is
the purpose of the present paper. To make progress on this issue, we report here the results of
probing the MGBC with Supernovae Ia (SNe Ia), direct observations of the Hubble parameter (OHD), Baryon Acoustic Oscillations (BAO) and Cosmic Microwave Background (CMB).
The agreement is very promising that allow us to conclude that the current accelerated expansion
might be of geometric origin. Moreover, based on these results, the idea that our $ 4D $ universe floats within a
fifth dimension must not be ruled out yet.
The inclusion of the $K$ term is not a simple improvisation. It has been considered in several
contexts ranging from the bending and shape determination of phospholipids membranes~\cite{Svetina1989},
in particle physics models to improve the earlier attempt by Dirac to picture the electron as a
bubble~\cite{Onder:1988wm,Cordero:2010vv}, in pure GR as being the well known Gibbons-Hawking-York
term, and recently as an effective $4D$ field yielding one of the Galileon actions with applications
in particle physics and cosmology~\cite{deRham:2010,Trodden2011}.




The paper is organized as follows. In Section II we briefly review the MGBC and
confine our attention to the expression for
the Hubble parameter in terms of the matter density parameters, that will be used in the
tests. In Section III the probes and statistical method are described. The results are discussed
in Section IV and finally the concluding remarks are given in the last Section.


\section{The Modified Geodetic Brane Cosmology}

Let us briefly review the MGBC model as well as the derivation of the
Friedmann equations that are taken as the basis for the cosmological data
confrontation. Detailed calculations can be found in~\cite{Cordero2012}.
Consider a spacelike 3-brane $\Sigma$, propagating in a flat
five-dimensional nondynamical Minkowski background spacetime with metric
$\eta_{A B}, \,(A, B =0,1,2,3,4)$.  The brane trajectory or worldvolume, $m$, is
an oriented timelike $4$-dimensional hypersurface parametrized by $y^A = X^A(x^\mu)$ in
the bulk where $y^A$ are the local coordinates for the background spacetime,
and $x^\mu$ are the local coordinates for $m$ while $X^A$ are the embedding functions
($\mu = 0, 1, 2, 3$). In general, any 4 dimensional brane Universe can be embedded at least locally for a background spacetime with dimension $N \geq 10$ \cite{Janet:1926,Cartan:1927}, and for $N \geq 91$  a global \cite{Clarke:1970} embedding is possible. In this spirit, a Friedmann-Robertson-Walker (FRW) metric can be embedded in a 5 dimensional background \cite{Rosen:1965}.

The dynamics of $\Sigma$ under the MGBC model is derived from the functional
\begin{equation}
S[X] = \int_{m} d^4 x \,\sqrt{-g}\left({\frac{\alpha}{2} {R}+ \beta K - \Lambda}\right),
\label{action}
\end{equation}
where the constants $\alpha$ and $\beta$ have the dimensions [L]$^{-2}$ and  [L]$^{-3}$ in Planck
units, respectively; $g = {\rm det} (g_{\mu \nu})$, $R$ stands for the worldvolume Ricci scalar,
$K = g^{\mu\nu}K_{\mu \nu}$ is the mean extrinsic curvature of $m$, and $\Lambda$ is the cosmological
constant defined on $m$. The action (\ref{action}) is invariant under reparametrizations of
the worldvolume and it leads to an only second-order equation of motion.

From the action it is obtained the generalized geodetic equation that governs the $\Sigma$ evolution,
\begin{equation}
(\alpha G^{\mu \nu}+ \beta S^{\mu\nu} + \Lambda g^{\mu\nu})K_{\mu\nu}=0,
\label{EqMotion}
\end{equation}
where $G_{\mu\nu}= R_{\mu\nu}- Rg_{\mu\nu}/2$ is the worldvolume Einstein tensor
and $S_{\mu\nu}:= K_{\mu\nu} -Kg_{\mu\nu}$. Notice that it is a second-order
differential equation written in a compact form. The tensors $G_{\mu\nu},S_{\mu\nu}$ are conserved in the
sense that $\nabla_{\mu} G^{\mu \nu}=0$ and $\nabla_{\mu} S^{\mu \nu}=0$ where
$\nabla_\mu$ is the covariant derivative compatible with the induced metric
$g_{\mu\nu}$ (see~\cite{CQG-2013} for more details).
We assume that the five-dimensional Minkowski spacetime, where $\Sigma$ evolves, is
\begin{equation}
ds_{5}^2= -dt^2+da^2+a^2 d\Omega_{3}^2,
\end{equation}
where $d\Omega_{3}^2$ denotes the unit 3-sphere. In addition, by assuming that on
large scales this type of universe is homogeneous and isotropic, the worldvolume
$m$ can be parametrized by $y^A=X^A(\xi^{\mu})=(t(\tau),a(\tau),\chi,\theta,\phi))$,
with $\tau$ being the proper time for an observer at rest w. r. t. the brane. This is
the usual Friedmann-Robertson-Walker (FRW) geometry. Thus,
the metric induced from the background spacetime is
\begin{equation}
ds_4^2=-N^2dt^2+a^2 d\Omega_{3}^2= g_{\mu\nu} dx^\mu dx^\nu.
\end{equation}
where $N= \sqrt{\dot{t}^2-\dot{a}^2}$. The evolution of the brane $\Sigma$, derived from
Eq.~(\ref{EqMotion}) under the FRW geometry is given by
\begin{equation}
\frac{d}{d \tau}\left({\frac{\dot{a}}{\dot{t}}}\right)= - \frac{N^2}{a^2(\dot{t}/a)}
\frac{(\dot{t}^2/a^2-3 \bar{\Lambda} N^2+6 \bar{\beta} N \dot{t}/a)}{(3 \dot{t}^2/a^2-
\bar{\Lambda} N^2+6 \bar{\beta} N \dot{t}/a)},
\end{equation}
where $\bar{x}$ denotes $x/(3 \alpha)$ and $\dot{x}$ means derivative w. r. t. the
parameter $\tau$.
Later on it will be clear that depending on its sign, the $\beta$ parameter contributes as
either a catalyst or a preventer of the acceleration of this type of universe.

So far this review comprises only the geometrical
information. If we intend to make contact with the cosmological implications of the model
we need to include the brane matter content which
should be compatible with the assumed symmetries. Thus, a perfect fluid fulfills these
requirements
\begin{equation}
T_{m}^{\mu\nu}=( \rho +P) \eta^\mu \eta^\nu + P g^{\mu\nu},
\end{equation}
where $\eta^\mu$ denotes the timelike unit normal vector to $\Sigma$, and $\rho (a)$ and
$P(a)$ are the energy density and pressure of the fluid, respectively.
It was also shown that since the model is invariant under reparametrizations in
the time coordinate $t$, the energy $E$ of the brane trajectory of this brane-like
universe results constant and given by
\begin{equation}
- \frac{E}{a^4} = \frac{(\dot{a}^2 + k)^{1/2}}{a} \left[ \frac{(\dot{a}^2 + k)}{a^2} -
(\overline{\Lambda} + \overline{\rho}) \right] + 3\overline{\beta}
\frac{(\dot{a}^2 + k)}{a^2},
\label{energy}
\end{equation}
where we have considered the three possible geometries for this type of universe
through the parameter $k$ ($k=-1,0,1$) besides of considering the so-called cosmic
gauge, $N=1$. Some remarks are in order. Expression (\ref{energy}) is nothing but the Friedmann equation associated
to the model (\ref{action}).
It is noteworthy that $E$, being and integration constant, it parametrizes the deviation from the Einstein limit
since as $E \to 0$ and $\beta \to 0$ together, the standard cosmology is recovered.
Finally, notice that this is a peculiar algebraic cubic equation in the combination
$(\dot{a}^2 + k)^{1/2}/a$ such that it only possesses a single physical solution \cite{Cordero2012}.

At this point it is important to remark the similarities and differences and between MGBC and UBC. The MGBC is one part of the UBC and it can be obtained if the gravitational effect of the brane on the bulk is not considered by means of taking the gravitational constant of the background spacetime very large (smooth background case).  The term $\frac{E}{a^4}$ was introduced before the RS radiation term and its origin is different because in the later case this term comes once the FRW brane is embedded in a Schwarzchild-AdS${}_5$ black hole background spacetime. Since the point of view of UBC (for $Z_2$ symmetry)  $E$ parametrizes the deviation from General Relativity and RS model and it is not necessary small.

Inclusion of matter will not affect the form of the equation of
motion~(\ref{EqMotion}). Hence, within this FRW geometry expression~(\ref{EqMotion}) in terms of the energy $E$ and in presence of a perfect fluid, reads
\begin{equation}
\frac{\ddot{a}}{a}= \frac{\left({\frac{\dot{t}^2}{a^2}}\right)\left[\frac{E}{a^4}
-3 \bar{\beta}\frac{\dot{t}^2}{a^2} +2 (\bar{\Lambda}+ \bar{\rho})\frac{\dot{t}}{a}-3(\bar{P}
+ \bar{\rho})\frac{\dot{t}}{a}\right]}{\left[-3 \frac{E}{a^4}-3 \bar{\beta}\frac{\dot{t}^2}{a^2}
+2 (\bar{\Lambda}+ \bar{\rho})\frac{\dot{t}}{a}\right]}.
\label{F_eq}
\end{equation}

A convenient form to rewritting the Friedmann equation (\ref{energy}) is in terms of the Hubble
parameter $H := \dot{a}/a$ as well as the energy density parameters associated with the model. Thus,
Eq. (\ref{energy}) becomes
\begin{align}
& \left({\frac{H^2}{H_0^2}- \frac{\Omega_{k,0}}{a^2}}\right)^{1/2} \left({\frac{H^2}{H_0^2} - \frac{\Omega_{k,0}}{a^2}-\frac{\Omega_{m,0}}{a^3}- \Omega_{\Lambda,0}}\right) + \nonumber \\
& \Omega_{\beta,0}\left({\frac{H^2}{H_0^2}- \frac{\Omega_{k,0}}{a^2}}\right) = \frac{\Omega_{dr}}{a^4},
\label{H_eq}
\end{align}

where the density parameters are defined by $\Omega_{\Lambda,0}:= \Lambda/(3 \alpha H_0^2)$,
$\Omega_{m,0}:= \rho_{m,0}/(3 \alpha H_0^2)$, $\Omega_{\beta,0}:= \beta/( \alpha H_0)$ and
the so called dark radiation density parameter $\Omega_{dr}:= -E/(H_0^3)$.
$H_0$ being the Hubble constant.

As it was pointed out in \cite{Cordero:2013pua}, the case when $\Omega_{dr}=0$ and
$\beta >0$ corresponds to the non-accelerating branch of the
DGP theory, whereas $\Omega_{dr}=0$ and $\beta <0$ conforms a self-accelerating branch
similar to the one of the DGP framework, and therefore appears a number of pathologies.
Notice also that the $\beta$ parameter plays the role of the inverse of the crossover
scale $r_{c}$, related to the transition from $4D$ to $5D$ in the DGP approach.
Furthermore, $\beta$ is related to the $5D$ Planck mass in the context of the
Galileon field theory of brane cosmology \cite{deRham:2010}.
The normalization condition is obtained by evaluating Eq. (\ref{H_eq}) at the present moment,
\begin{equation}
\left({1- \Omega_{k,0}}\right)^{1/2} \left({1-\Omega_{k,0}-\Omega_{m,0}- \Omega_{\Lambda,0}}\right)
+ \Omega_{\beta,0} \left({1- \Omega_{k,0}}\right) = \Omega_{dr}.
\label{Eq:normalization}
\end{equation}
As before, when $\Omega_{dr}=0$ and $\Omega_{\beta,0}=0$ the standard cosmology
is recovered, accordingly. The general behaviour of the scale factor is studied in detail in~\cite{Cordero2012}. Now, (\ref{H_eq}) is an algebraic cubic equation in $H^2/H_0^2$
whose real root has been determined. This shall be used to compare with the cosmological data
in the next Section. For sake of illustration, in the analysis below we shall assume
a flat FRW type universe, i.e., by considering $\Omega_{k,0}=0$



\section{Statistics and Observational Samples}

\subsection{Statistical Method}

To find the high confidence region of the parameter space of the MGBC given a set of observational
data, we use the Markov Chain Monte Carlo (MCMC) method, which is an algorithm with the capability
of providing robust and realistic constraints on cosmological model parameters. The method is fairly
standard but suffers of an important complication consisting in how to approach correctly the
convergence of the chain. In particular, our code addresses this issue following the prescription
developed in \cite{Dunkley05}. See \cite{Berg,MacKay,Neal} and references therein, for a review of
MCMC methods.

By using our MCMC code, we perform a likelihood analysis in which we minimize the $\chi^2$ function
thus obtaining the best fit of model parameters from  observational data. This minimization is
equivalent to maximizing the likelihood function $\mathcal{L}(\theta) \propto \exp [-\chi^2(\theta)/2]$
where $\theta$ is the vector of model parameters. The expression for $\chi^2(\theta)$ depends on the
used dataset. In what follows we briefly describe the probes and samples.




\subsection{Type Ia Supernovae (SNe Ia)}

We consider the Union2.1 compilation of 580 SNe Ia in the redshift range $0.015<z< 1.414$ reported by the Supernova Cosmology
Project (SCP) \cite{Union21}. As is usual, the comparison with SNe Ia data is made via the
standard $\chi^2$ statistics given by
\begin{equation}
\chi^2_{SNe Ia}= \mathrm{\Delta} \mathrm{\mu} \cdot \mathrm{C^{-1}} \cdot \mathrm{\Delta} \mathrm{\mu},
\label{Eq:CSNIa}
\end{equation}
where $\mathrm{C}$ is the covariance matrix and $\mathrm{\Delta} \mathrm{\mu}=\mathrm{\mu_{th}}
-\mathrm{\mu_{obs}}$ is the vector of the differences between the observed and theoretical value
of the distance modulus $\mathrm{\mu}$. For Union2.1, $\mathrm{C}$ captures all identified
systematic errors besides to the statistic errors of the SNe Ia data and $\mathrm{\mu}$ is defined by
\begin{equation}
\mu(z,\theta)= 5 \log_{10} \left[d_L(z,\theta) \right] + \mu_0,
\label{Eq:muSN}
\end{equation}
where $d_L(z,\theta) $ is the dimensionless luminosity distance given by
\begin{equation}
d_L(z,\theta)= (1+z) \int_0^z \frac{dz'}{E(z',\theta)},
\end{equation}
with $E(z,\theta)=H(z,\theta)/H_0$ the dimensionless Hubble function, $H_0$ the Hubble constant
and $\theta$ the free parameters of the cosmological model.

In Eq. (\ref{Eq:muSN}) $\mu_0$ is a nuisance parameter that depends on both the absolute magnitude
of a fiducial SN Ia and the Hubble constant. Here, we marginalize the $\chi^2_{SNe Ia}$ over $\mu_0$.

\subsection{Observational Hubble Data (OHD)}

The observational Hubble parameter $H(z)$ data provide a direct measurement of the Hubble parameter,
instead of its integral like the SNe Ia or the Baryon Acoustic Oscillations (BAO) probes.

So far the main complication of the OHD data is the number of data points available in comparison
with SNe Ia luminosity distance data. However, we think the OHD data can help break the parameter
degeneracies and shed light on the cosmological scenarios that are being studied.

In this work, we use 18 data points from differential evolution of passively evolving early-type
galaxies in the redshift range $0<z<1.75$ reported in \cite{Jimenez12}.

The best fit values of the model parameters from OHD are determined by minimizing the quantity
\begin{equation}
\chi^2_{OHD}= \sum^{18}_{j=1} \frac{\left[ H_{th}(z_j,\mathbf{\theta})
-H_{obs}(z_j)\right]^2}{\sigma^2_{H_{obs}}(z_j)},
\label{Eq:COHD}
\end{equation}
where $\sigma^2_{H_{obs}}$ are the measurement variances, and $\mathbf{\theta}$ corresponds to the free
parameters of the cosmological model.

\subsection{Baryon Acoustic Oscillations (BAO)}

Currently BAO is considered a promising standard ruler to use in cosmology enabling precise measurements of the distance ratio, $D_V /r_s$, the distance to objects at redshift $z$ in units of the sound horizon at recombination, independently of the local Hubble constant. The observed angular and radial BAO scales at redshift $z$ provide a geometric estimate of the effective distance,
\begin{equation}
D_V(z) \equiv \left[(1+z)^2D_A^2(z) cz/H(z)\right]^{1/3},
\end{equation}
where $D_A(z)$ is the angular diameter distance and $H(z)$ is the Hubble parameter. The measured ratio $D_V/r_s$, with $r_s$ being the comoving sound horizon scale at the end of the drag epoch, is what can be compared to theoretical predictions.

Since the release of the seven-year WMAP data, the acoustic scale measurement has been improved by the Sloan Digital Sky Survey (SDSS) and SDSS-III Baryon Oscillation Spectroscopic Survey (BOSS) galaxy surveys, and by the WiggleZ and 6dFGS surveys. An upgraded estimate of the acoustic scale in the SDSS-DR7 data was made in \cite{Padmanabhan2012MNRAS}, giving $D_V(0.35)/r_s=8.88\pm 0.17$, and reducing the uncertainty from 3.5$\%$ to 1.9$\%$. More recently, the SDSS-DR9 data from the BOSS survey has been used to estimate the BAO scale of the CMASS sample. In this last case, in \cite{Anderson2012MNRAS} was reported $D_V(0.57)/r_s=13.67\pm 0.22$ for galaxies in the range $0.43 < z < 0.7$ (at an effective redshift $z= 0.57$). On the other hand, the acoustic scale has also been measured at higher redshift using the WiggleZ galaxy survey. In \cite{Blake:2012pj} was reported distances in three correlated redshift bins between $0.44$ and $0.73$. At lower redshift, $z = 0.1$, a detection of the BAO scale has been made using the 6dFGS survey, see \cite{Beutler:2011hx}. To perform our analysis we employed the data described before; a summary of these measurements can be found in Table 1 of \cite{Hinshaw2013}.

The corresponding $\chi^2$ is given by
\begin{equation}
\chi^2_{BAO}=(v_i-v_i^{BAO})(\mathbf{C}^{-1})_{ij}^{BAO}(v_j-v_j^{BAO})^T,
\label{Eq:chiBAO}
\end{equation}
where $(\mathbf{C}^{-1})^{BAO}$ is the inverse covariance matrix of the data and $v_i$ is given by $v_i=\left\lbrace \frac{r_s(z_{drag};\mathbf{\theta})}{D_V(z;\theta)} \right\rbrace$.

\subsection{Cosmic Microwave Background (CMB)}

We also use the CMB data reported in \cite{Wang:2013mha}, where distance priors are derived from the first release of the Planck results.

CMB data provide the comoving distance to the photon-decopling surface $r(z_{*})$, $z_*=1090$, and the comoving sound horizon at photon-decopling epoch $r_s(z_{*})$ which are defined as,
\begin{equation}
r(z_*)=\frac{c}{H_0} \int_0^{z_*} \frac{dz'}{E(z')},
\end{equation}
and
\begin{eqnarray}
r_s(z_*)&=&\frac{c}{H_0} \int_{z_*}^{\infty} \frac{c_s}{E(z')} dz' \\ \nonumber
           &\equiv & \frac{c}{H_0} \int_0^{a_*}\frac{da'}{\sqrt{3(1+\overline{R_b} a')a'^4 E^2(a')}},
\end{eqnarray}
where $\overline{R_b} a=3\rho_b/(4\rho_{\gamma})=31500\Omega_bh^2(T_{CMB}/2.7K)^{-4}a$.

The CMB shift parameters are
\begin{equation}
R \equiv \sqrt{\Omega_m H_0^2} \frac{r(z_{*})}{c},
\end{equation}
\begin{equation}
l_a\equiv \pi \frac{r(z_{*})}{r_s(z_{*})},
\end{equation}
and both are evaluated at photon-decoupling epoch $z_*$. The baryon density $\Omega_bh^2$ is also considered as shift parameter.

The mean values for these shift parameters as well as their standard deviations and normalized covariance matrix are taken from \cite{Wang:2013mha}. Thus, the corresponding $\chi^2$ for the CMB is
\begin{equation}
\chi^2_{CMB}=(v_i-v_i^{CMB})(\mathbf{C}^{-1})_{ij}^{CMB}(v_j-v_j^{CMB})^T,
\end{equation}
where $(\mathbf{C}^{-1})^{CMB}$ is the inverse covariance matrix of the data and $v_j$ means the vector parameter $(\mathbf{v})_j$,  $\mathbf{v}=(l_a,R,\Omega_bh^2)$.


\section{Observational results}

\begin{table*}[!htp]
\caption{Summary of the Modified GBG cosmological scenarios tested.}
\renewcommand{\arraystretch}{2.}
\begin{tabular}{l @{\extracolsep{10pt}} l @{\extracolsep{15pt}} l }
\hline
Case & Free Parameters & Assumptions \\ \hline
\vspace{2mm}
\multirow{1}{*}{I} & $\Omega_{\beta}$, $\Omega_{dr}$, $\Omega_{\Lambda}$  & $\Omega_m=0.315$ \\ \hline
 \vspace{2mm}
\multirow{1}{*}{II} & $\Omega_{\beta}$, $\Omega_{dr}$, $\Omega_m$ & $\Omega_{\Lambda}=0$  \\ \hline
 \vspace{2mm}
\multirow{1}{*}{III} & $\Omega_{\beta}$, $\Omega_{m}$, $\Omega_{\Lambda}$  & $\Omega_{dr}=\Omega_r$\\ \hline
\vspace{2mm}
\multirow{1}{*}{IV} & $\Omega_{\beta}$, $\Omega_{m}$ & $\Omega_{dr}=\Omega_r$, $\Omega_{\Lambda}=0$\\ \hline
\vspace{2mm}
\multirow{1}{*}{V} & $\Omega_{\beta}$, $\Omega_{dr}$ & $\Omega_m=0.315$, $\Omega_{\Lambda}=0$\\ \hline

\end{tabular}
\label{Table:Models}
\end{table*}

In this section we will discuss the results of probing the MGBC
cosmology with the observational samples. Five cases will be considered, some of them
being very promising in explaining the current accelerated expansion.

In Table \ref{Table:Models}, we present a summary of the cases analyzed. \textit{Case I} has as free
parameters $\Omega_{\beta}$, $\Omega_{dr}$ and $\Omega_{\Lambda}$; for this case it is assumed
the matter density of the universe is $\Omega_m=0.315$ as the Planck's results suggest \cite{PlanckXVI}.
\textit{Case II} has $\Omega_{\beta}$, $\Omega_{dr}$ and $\Omega_m$ as free parameters and we consider
$\Omega_{\Lambda}=0$. \textit{Case III} consists in taking $\Omega_{\beta}$, $\Omega_{m}$,
$\Omega_{\Lambda}$ as free parameters and $\Omega_{dr}$ corresponding to the radiation density,
$\Omega_r$. \textit{Case IV} has as free parameters just to $\Omega_{\beta}$, $\Omega_{m}$ and it
is assumed that $\Omega_{dr}=\Omega_r$, $\Omega_{\Lambda}=0$. \textit{Case V} consists in
taking $\Omega_{\beta}$, $\Omega_{dr}$ as free parameters of the model with the assumptions
of $\Omega_m=0.315$, $\Omega_{\Lambda}=0$.

\begin{table}[!htbp]
\caption{Summary of the best estimates of model parameters for the Case I as well as for the best values for
the  parameter $\Omega_{\Lambda}$ coming from the normalization condition, Eq. (\ref{Eq:normalization}) with $\Omega_{k,0}=0$. The errors are at $68.3\%$ confidence level. Here, to obtain the  best estimates of model parameters, it has been
assumed a prior on $\Omega_m=0.315\pm0.012$ from the first Planck results \cite{PlanckXVI}.}
\label{Table1}
\begin{tabular}{|c||c||c||c||}
\hline
 & \multicolumn{1}{c||}{SNIa}&  \multicolumn{1}{c||}{OHD}    & \multicolumn{1}{c||}{Joint} \\
\hline
$\Omega_{\beta}$ & $-0.074^{+0.191}_{-0.254}$&  $-0.136^{+0.347}_{-0.473}$
& $-0.084^{+0.204}_{-0.276}$  \\

$\Omega_{dr}$ & $0.028^{+0.031}_{-0.028}$  & $0.038^{+0.068}_{-0.038}$
& $0.023^{+0.038}_{-0.023}$  \\

$\chi^2_r$& 0.959 &  0.989 & 0.952 \\
\hline
$\Omega_{\Lambda}$ &$0.580^{+0.163}_{-0.223}$ &$0.507^{+0.283}_{-0.431}$  &$0.577^{+0.168}_{-0.252}$
\\

\hline
\hline
\end{tabular}
\end{table}

Firstly we will describe the results obtained with SNIa and OHD data, while in the next subsection the BAO and CMB probes will be incorporated to the analysis.

The best fit parameter values with 1$\sigma$ error and the corresponding values of $\chi^2$ obtained by using SNIa data, OHD data and the joint of these data sets are summarized case by case in Tables \ref{Table1} to \ref{Table5}.

Table \ref{Table1} shows the results for the Case I. In this scenario the  preferred amounts for $\Omega_{\beta}$ and $\Omega_{dr}$ from both data sets, SNIa data and OHD data, are very small; however by combining these data sets, the best-fit parameter
values increase but not significantly. This results suggest that $\Omega_{\Lambda}$, although smaller than the one in $\Lambda$CDM,
has to be present in order to drive the accelerated expansion.


For the Case II, the best-fit for $\Omega_{\beta}$ resulted to be as the MGBC model dictates in order to produce the cosmic
acceleration, a negative quantity. Furthermore, $\Omega_{\beta}$ contributes in a
significantly way. However from the normalization condition, Eq. (\ref{Eq:normalization}), it is found that the amount of
matter, $\Omega_m$, is inconsistent with the result given by  Planck's \cite{PlanckXVI}, see Table \ref{Table2}. This case is therefore ruled out for explaining the accelerated expansion.



\begin{table}[!htbp]
\caption{Best estimates of model parameters for the Case II as well as for the best values for
the  parameter $\Omega_{m}$ from the normalization condition. The errors are at $68.3\%$
confidence level. }
\label{Table2}
\begin{tabular}{|c||c||c||c||}
\hline
 & \multicolumn{1}{c||}{SNIa}&  \multicolumn{1}{c||}{OHD}    & \multicolumn{1}{c||}{Joint} \\
\hline
$\Omega_{\beta}$ & $-0.808^{+0.056}_{-0.077}$ & $-0.745^{+0.050}_{-0.073}$& $-0.779^{+0.046}_{-0.055}$ \\

$\Omega_{dr}$ & $0.089^{+0.127}_{-0.089}$ &$0.127^{+0.163}_{-0.127}$ & $0.103^{+0.131}_{-0.103}$ \\

$\chi^2_r$&0.955&  0.986  & 0.951 \\
\hline
$\Omega_{m}$      &$0.099^{+0.149}_{-0.099}$ &$0.125^{+0.177}_{-0.125}$ &  $0.116^{+0.149}_{-0.116}$   \\
\hline
\hline
\end{tabular}
\end{table}

Table \ref{Table3} summarizes the results for the Case III. For this scenario,  assuming $\Omega_{dr}$ equal to $\Omega_{r}$, the obtained best-fit
value for $\Omega_{\beta}$  is almost negligible,
meaning that the model cannot explain the accelerated expansion through this parameter,
and instead $\Omega_{\Lambda}$  dominates the Universe in almost the same quantity dictated
by the $\Lambda$CDM model.

Table \ref{Table4} shows the observational constraints for the Case IV. It is assumed $\Omega_{\Lambda}=0$ and $\Omega_{dr}=\Omega_r$. The results coming from the OHD data, the SNIa data and the combination of the two samples point out that $\Omega_{\beta}$ could be responsible of cosmic acceleration.
On the other hand, from the normalization condition is found that the matter content is smaller than the
accepted value.

\begin{table}[!htp]
\caption{Summary of the best estimates of model parameters for Case III as well as the best values for the
parameter $\Omega_{\Lambda}$ obtained by the normalization condition. The errors are at $68.3\%$
confidence level. To obtain the  best estimates it has been assumed
$\Omega_{dr}=\Omega_{r}=\Omega_{\gamma}(1+0.2271 N_{eff})$.}
\label{Table3}
\begin{tabular}{|c||c||c||c||}
\hline
 & \multicolumn{1}{c||}{SNIa}&  \multicolumn{1}{c||}{OHD}    & \multicolumn{1}{c||}{Joint} \\
\hline
$\Omega_{\beta}$ & $0.028^{+0.759}_{-0.869}$& $0.007^{+0.793}_{-0.791}$
& $0.073^{+0.703}_{-0.868}$  \\
$\Omega_{m}$ & $0.300^{+0.174}_{-0.145}$  & $0.343^{+0.145}_{-0.131}$ & $0.331^{+0.133}_{-0.128}$
 \\
$\chi^2_r$& 0.955 &  0.831 &0.949  \\
\hline
$\Omega_{\Lambda}$ &$0.722^{+0.593}_{-0.722}$ &$0.661^{+0.654}_{-0.661}$  &$0.740^{+0.575}_{-0.740}$
\\
\hline
\hline
\end{tabular}
\end{table}

\begin{table}[!htp]
\caption{Best estimates of model parameters for Case IV as well as the best values for the
parameter $\Omega_{m}$ obtained from the normalization condition. The errors are at $68.3\%$
confidence level. To obtain the  best estimates it has been assumed
$\Omega_{dr}=\Omega_{r}=\Omega_{\gamma}(1+0.2271 N_{eff})$.}
\label{Table4}
\begin{tabular}{|c||c||c||c||}
\hline
 & \multicolumn{1}{c||}{SNIa}&  \multicolumn{1}{c||}{OHD}    & \multicolumn{1}{c||}{Joint} \\
\hline
$\Omega_{\beta}$ &$-0.797^{+0.037}_{-0.031}$ &$-0.742^{+0.047}_{-0.048}$& $-0.768^{+0.035}_{-0.035}$ \\
$\chi^2_r$ & 0.953 &0.833 & 0.948 \\
\hline
$\Omega_{m}$      &$0.203^{+0.037}_{-0.031}$ &$0.258^{+0.047}_{-0.048}$ &$0.232^{+0.035}_{-0.035}$   \\
\hline
\hline
\end{tabular}
\end{table}

Finally, Table \ref{Table5} shows the results for the scenario (Case V) in which $\Omega_{\Lambda}=0$ and $\Omega_m=0.315$. The free parameter of the model is $\Omega_{\beta}$,  and from the normalization
condition straightforwardly is inferred $\Omega_{dr}$. In this case pretty
good restrictions are obtained. The amount predicted  for $\Omega_{\beta}$ is negative and very close to the abundance ascribed to $\Omega_{\Lambda}$. Taking together SNIa and OHD data we
obtained as best-fit model parameter $\Omega_{\beta}=-0.679^{+0.002}_{-0.006}$ and
$\Omega_{dr}=0.007^{+0.003}_{-0.007}$, from the normalization condition. As far as these probes concern, this is a good candidate to account for the observed accelerated expansion.

\begin{table}[!htp]
\caption{Summary of the best estimates of model parameters as well as for the best values for
the  parameter $\Omega_{dr}$. The errors are at $68.3\%$ confidence level. In this case
$\Omega_{\Lambda}=0$ and $\Omega_m=0.315$ (Case V).}
\label{Table5}
\begin{tabular}{|c||c||c||c||}
\hline
 & \multicolumn{1}{c||}{SNIa}&  \multicolumn{1}{c||}{OHD}    & \multicolumn{1}{c||}{Joint} \\
\hline

$\Omega_{\beta}$ & $-0.673^{+0.005}_{-0.012}$& $-0.673^{+0.008}_{-0.012}$& $-0.679^{+0.002}_{-0.006}$  \\
$\chi^2_r$& 0.972 & 1.069 & 0.969 \\
\hline
$\Omega_{dr}$ &$0.012^{+0.005}_{-0.012}$ & $0.012^{+0.008}_{-0.012}$ & $0.007^{+0.003}_{-0.007}$ \\

\hline
\hline
\end{tabular}

\end{table}

\subsection{Complementing  with BAO and CMB probes}

Up to now we have performed observational tests by using SNIa and H(z) data, thus we have only focused on low-redshift probes. To arrive to more robust conclusions it is worth to study the MGBC model at high redshifts. In order to do that, we add a term of standard radiation,  $\Omega_r$, in the Hubble parameter equation, Eq. (\ref{H_eq}),  then the model becomes suitable to be confronted with BAO and CMB. BAO data is considered a powerful probe of dark energy and in turn, CMB data provide the strongest constraints on cosmological parameters thus being capable to break degeneracies between dark energy and cosmological parameters.


In Table \ref{Table6} we present the results of the inclusion of these data sets. The constraints were derived from the union of all the observational data sets: SNIa, $H(z)$, BAO and CMB; in all cases we assume $\Omega_r=\Omega_{\gamma}(1+0.2271N_{eff})$ from the Planck results. By comparing the new results with the previous ones (SNIa and $H(z)$ data), we found a good agreement between them, thus validating our earlier conclusions. Notice that an important difference with this inclusion is that tighter constraints are obtained, which are easily observed in the errors of the cosmological parameters.

\begin{table*}[!htbp]
\caption{Summary of the best estimates of model parameters obtained by the joint of SNIa, $H(z)$, BAO and CMB data for all the scenarios studied before. Separated of the former by a horizontal line are the  parameters obtained from their respective normalization condition, $\Omega_{\Lambda}=0$, \textit{prior} means $\Omega_m= 0.315$ was assumed. For Cases III and IV, $\Omega_{dr}=\Omega_r$. The errors are at $68.3\%$ confidence level. }
\label{Table6}
\begin{tabular}{|c||c||c||c||c||c||}
\hline
& \multicolumn{1}{c||}{Case I}&  \multicolumn{1}{c||}{Case II}    & \multicolumn{1}{c||}{Case III}  & \multicolumn{1}{c||}{Case IV} & \multicolumn{1}{c||}{Case V}\\
\hline
$\Omega_{\beta}$ & $-0.023^{+0.035}_{-0.063}$&$-0.759\pm 0.004$ &  $-0.119^{+0.097}_{-0.099}$
&$-0.759\pm0.004$& $-0.6848\pm0.0001$ \\

$\Omega_{dr}$ & $0.002^{+0.004}_{-0.002}$  &$0.0004^{+0.0001}_{-0.0004}$ & --
&-- &--  \\
$\Omega_m$ & --  &-- & $0.304\pm0.012$
&-- &--  \\
$\chi^2_r$& 0.946 & 1.118& 0.943 &1.119 & 1.853  \\
\hline
$\Omega_{\Lambda}$ &$0.660^{+0.032}_{-0.060}$ & 0 &$0.577^{+0.087}_{-0.089}$  & 0 & 0
\\
$\Omega_{m}$      & \textit{prior} &$0.241^{+0.005}_{-0.003}$ &--  &$0.241\pm 0.004$ & \textit{prior}
  \\
$\Omega_{dr}$      & -- &-- &$\Omega_r$  & $\Omega_r$& $(8.9^{+3.5}_{-8.9})\times 10^{-5}$
  \\
\hline
\hline
\end{tabular}
\end{table*}

\begin{figure}
  \centering
  \includegraphics[width=0.4\textwidth]{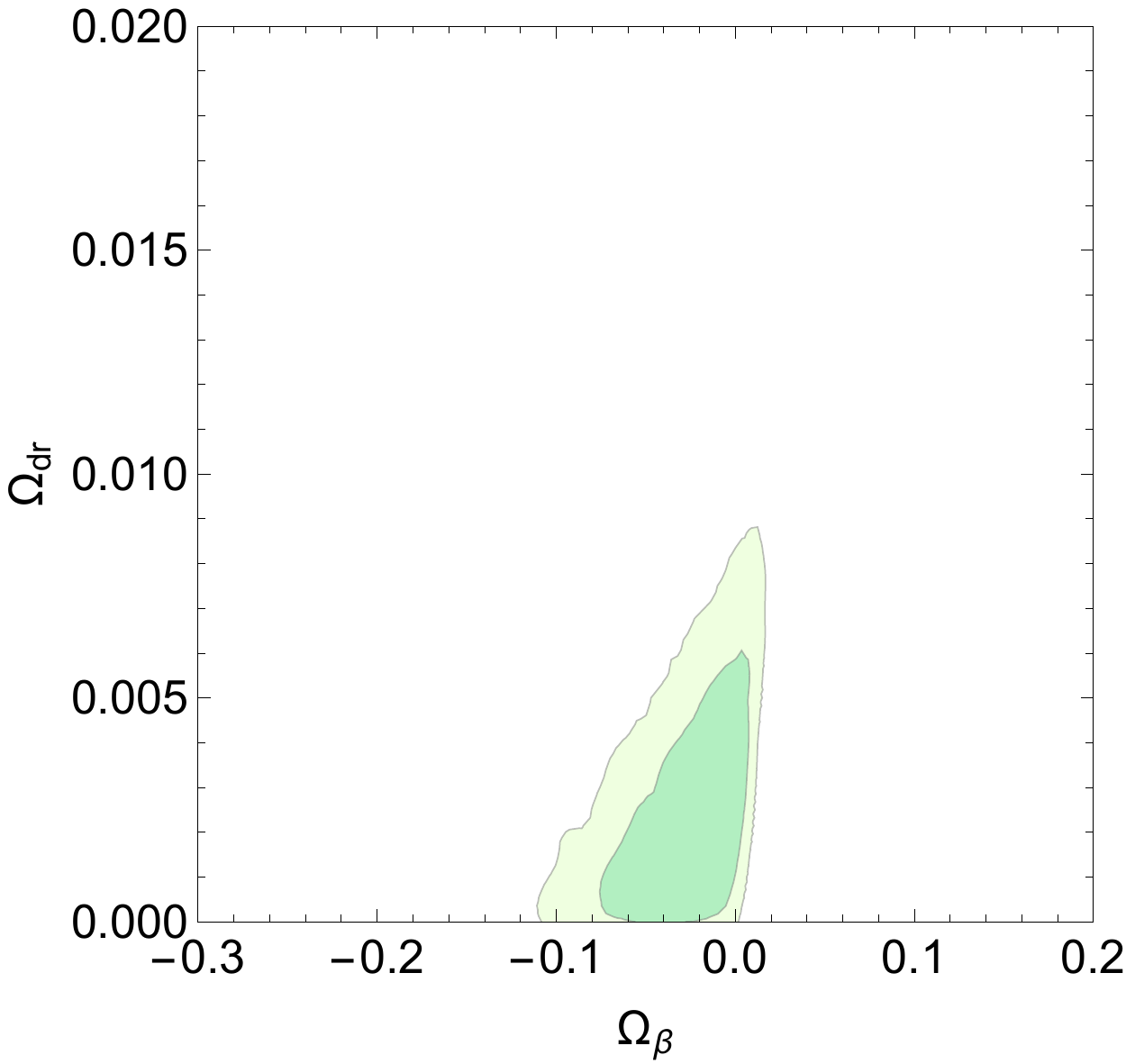}
  \hspace{5mm}
  \includegraphics[width=0.4\textwidth]{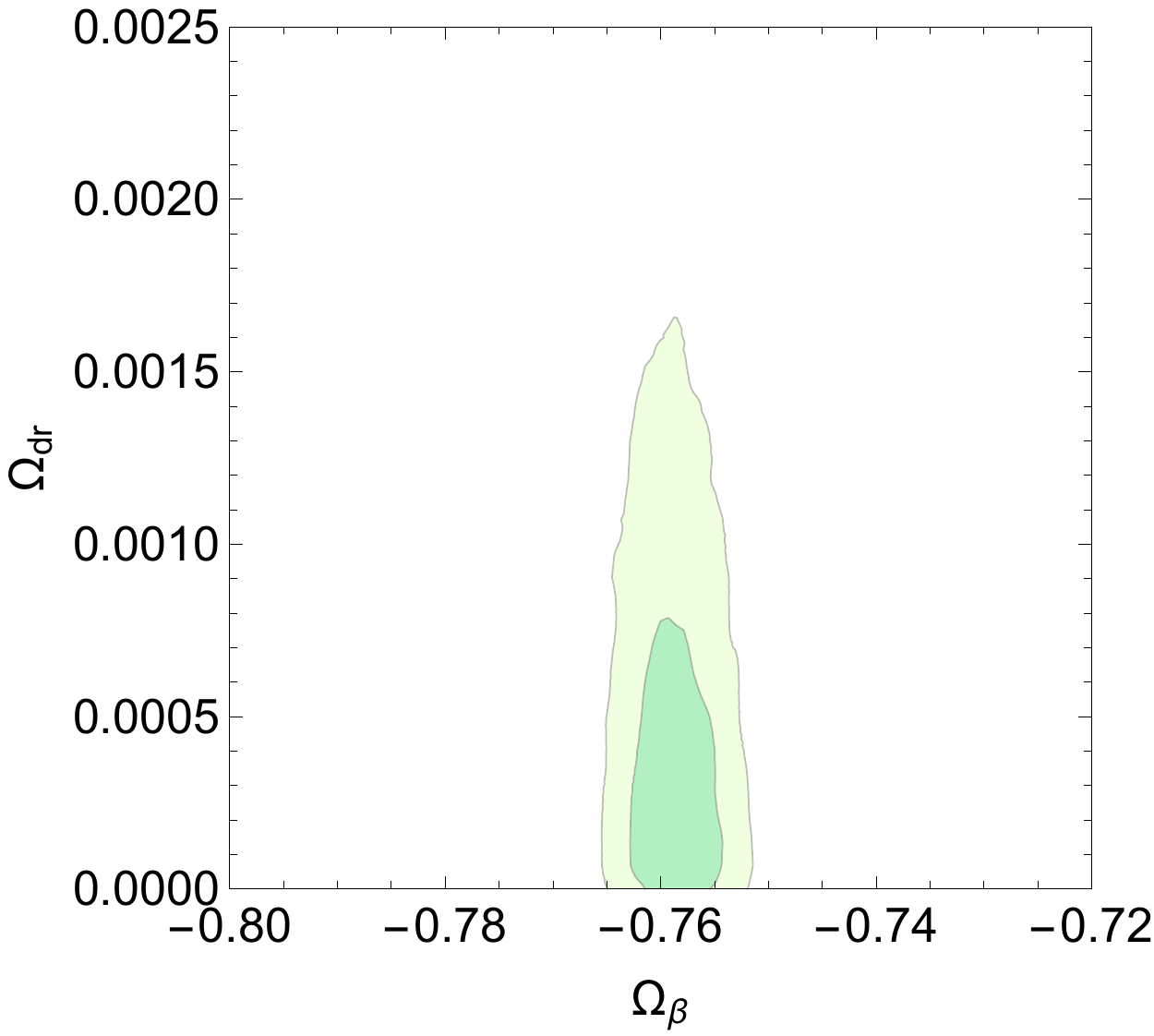}
  \caption{1$\sigma$ and 2$\sigma$ contours in the $\Omega_{\beta}$-$\Omega_{dr}$
  parameter space. Contours from the joint of SNIa, $H(z)$, BAO and CMB data.   CASE I (\textit{Top Panel}): It has been considered a prior on $\Omega_m$
  from the Planck results \cite{PlanckXVI}. CASE II (\textit{Bottom Panel}) : For this case no prior has been assumed on $\Omega_m$. For both cases it has been assumed $\Omega_r=\Omega_{\gamma}(1+0.2271N_{eff})$}
  \label{Figure1}
\end{figure}

\begin{figure}
  \centering
  \includegraphics[width=0.4\textwidth]{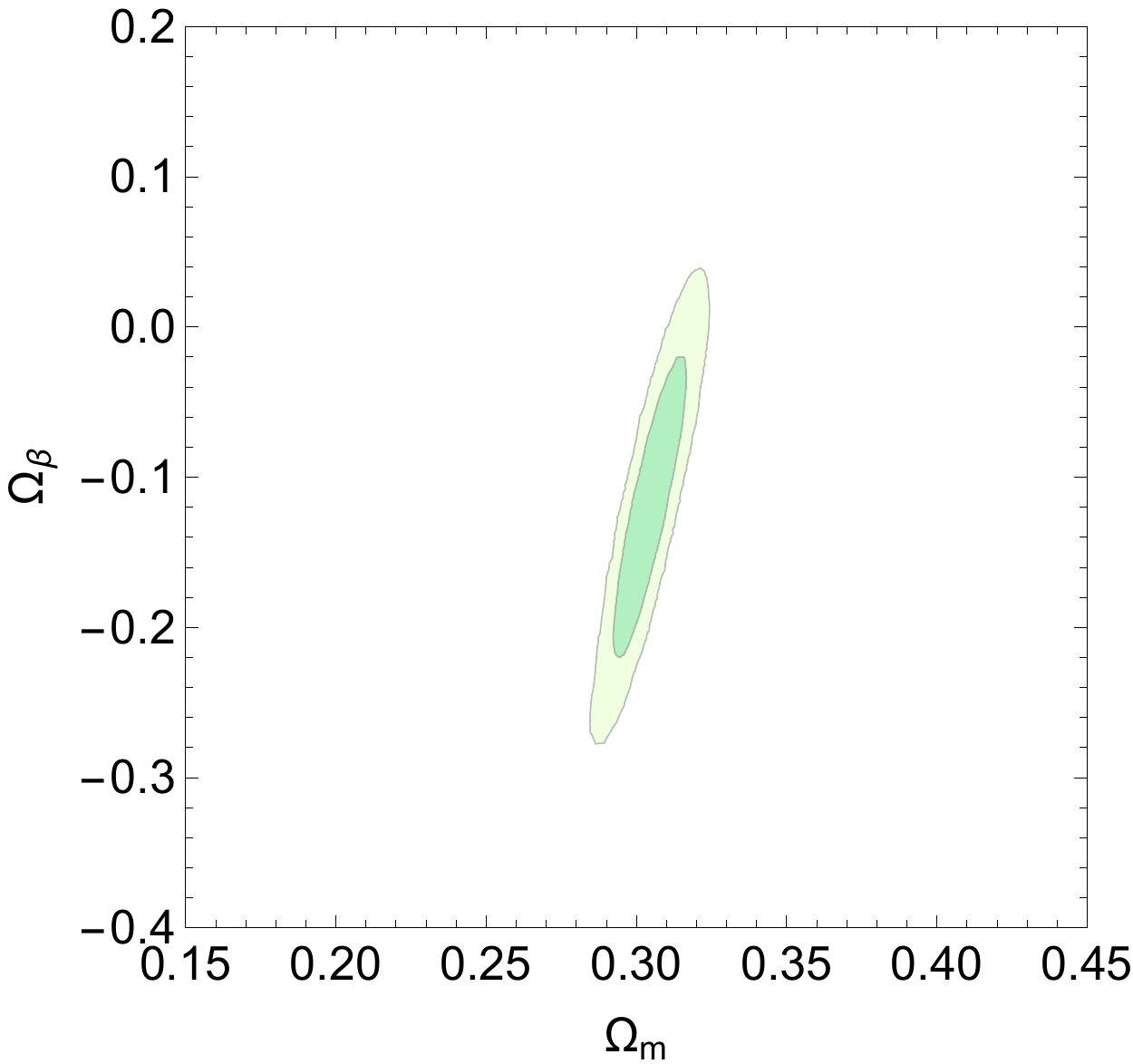}
  \caption{CASE III: 1$\sigma$ and 2$\sigma$ contours in the $\Omega_{m}$-$\Omega_{\beta}$
  parameter space. Contours come from the joint of SNIa, $H(z)$, BAO and CMB data. }
  \label{Figure2}
\end{figure}

In  Figure \ref{Figure1}, results are shown for Cases I and II;  both panels show contours in the $\Omega_{\beta}$-$\Omega_{dr}$ parameter space.  On the other hand, Figure \ref{Figure2} shows the constraints on the free parameters ($\Omega_m$ , $\Omega_{\beta}$) for the third scenario. In both figures 68$\%$ and 95$\%$ confidence level contours are displayed.

The Case V deserves special attention. Notice that from the combination of all data sets, the amount obtained of $\Omega_{\beta}$ suggests that it could account for cosmic acceleration. It is well known that the reduced $\chi^2$ gives us an idea of the quality of the fit, which should be close to 1 to be considered good. In this case the obtained $\chi^2_r=1.853$ is a little bit high. With this in mind we examined once again the scenario but taking out the CMB data. Then from the combined probes SNIa, OHD and BAO the obtained adjustment is  $\Omega_{\beta}= -0.682 ^{+0.001}_{-0.003}$ with a $\chi^2_r=0.982$; from the normalization condition the best value for $\Omega_{dr}$ is $\Omega_{dr}=0.003^{+0.001}_{-0.003}$. These results are in excellent agreement with previous tests and additionally the fit is very good.

Finally, with the goal of getting more robust conclusions about the power of the MGBC model in explaining the current accelerated expansion, we perform an analysis by keeping free all the cosmological parameters, namely, $\Omega_{\beta}$, $\Omega_{dr}$, $\Omega_m$, $\Omega_r$ and $h$. With the normalization condition, the value for $\Omega_{\Lambda}$ can be obtained with $\Omega_{\Lambda}=1+\Omega_{\beta}-\Omega_{dr}-\Omega_m-\Omega_r $. By combining all the used data sets  we get $\Omega_{\beta}=-0.024^{+0.112}_{-0.082}$, $\Omega_{dr}=0.007^{+0.012}_{-0.007}$, $\Omega_{m}=0.301^{+0.014}_{-0.019}$, $\Omega_r=[9.157^{+0.307}_{-0.652}]\times 10^{-5}$ and $h=0.690^{+0.035}_{-0.022}$ with $\chi^2_r=0.945$. One can immediately note that the fit is pretty good, but in cases with a non zero cosmological constant, $\Omega_{\beta}$ is not enough to produce the current cosmic acceleration, instead $\Omega_{\Lambda}$ contributes to the total of matter content with $\Omega_{\Lambda}=0.667^{+0.132}_{-0.082}$, in agreement with Planck results \cite{PlanckXVI}.

On the other hand, in the cases with $\Omega_{\Lambda}=0$ (cases II, IV, V), from the combined analysis including the four probes (see table \ref{Table6}), it turned out that $\Omega_{\beta}$ is able to account for the accelerated expansion, and so the three of them, specially case V, are good candidates in favor of considering a geometrical origin of the accelerated expansion.



\section{Concluding Remarks}

In this work it is examined how well the cosmological observations fit with the
parameters of the modified geodetic brane cosmology (MGBC). Four probes have been considered:
the supernovae Ia observations; the direct measurements of the Hubble parameter
from differential evolution of passively evolving early-type galaxies;  measurements of the distance ratio, $D_V /r_s$ gained from BAO measurements and finally, measurements of CMB shift parameters which provide an efficient summary of CMB data.

The set of parameters to be tested are $\theta_{i}=(\Omega_{m,0},
\Omega_{\Lambda,0}, \Omega_{\beta}, \Omega_{dr})$, taken in pairs, fixing one of them
and determining the fourth using the consistency relation; five cases were considered (see Table \ref{Table:Models}).
For the first one were considered the following parameters I:$\theta_{i}=(\Omega_{\beta,0},
\Omega_{dr}, \Omega_{\Lambda,0})$ assuming the value of $\Omega_m=0.315\pm0.012$ from the
first Planck results \cite{PlanckXVI}. It turns out that this case is unable to explain the cosmic
acceleration without cosmological constant. The second case consist of the specifications
II:$\theta_{i}=(\Omega_{\beta,0}, \Omega_{dr}, \Omega_{m,0})$ without cosmological constant
$\Omega_{\Lambda,0}=0$; with this setting the right cosmological acceleration has its
origin in the $\beta$ parameter; however the resulting matter content is a little bit smaller than
the accepted value. The third case is proposed as III:$\theta_{i}=(\Omega_{\beta,0},
\Omega_{m,0}, \Omega_{\Lambda,0})$; $\Omega_{dr}$ is assumed to be the CMB radiation;
the contribution of $\Omega_{\beta,0}$ is not enough to accelerate and cosmological constant
is needed; the resulting matter content is correct; this case resembles pretty much the $\Lambda$CDM model.
Case IV:$\theta_{i}=(\Omega_{\beta,0}, \Omega_{m,0})$ without cosmological
constant $\Omega_{\Lambda,0}=0$ and considering that $\Omega_{dr}$ is the CMB radiation;
$\Omega_{\beta,0}$ accelerates the universe in the right amount but the matter content value is
below to the accepted one.

The last and most promising case V is:$\theta_{i}=(\Omega_{\beta,0},
\Omega_{dr})$ without cosmological constant $\Omega_{\Lambda,0}=0$ and assuming
$\Omega_m=0.315\pm0.012$; $\Omega_{\beta,0}$ accelerates the universe in the right amount.
The best fit is obtained for this set of parameters with the joint probes SNIa, OHD and BAO, for redshifts in the range
$0<z< 1.75$. See Table \ref{Table6} for a comparison
of the adjustments of the five cases.

Finally it is remarkable that in cases with $\Omega_{\Lambda}=0$ (II, IV, V), the adjustment of $\Omega_{\beta}$ significantly increases its absolute value, supporting the argument that indeed $\Omega_{\beta}$ can play the role of a cosmological constant. The obtained results confirm that
the currently observed accelerated expansion might have a geometrical origin, since the
introduced parameter corresponding to the extrinsic curvature of the brane, is able to
produce the observed acceleration.

\begin{acknowledgments}
A.M.  acknowledges financial support by Conacyt through a Ph. D. fellowship;
R.C. acknowledges financial support by SNI, EDI, COFAA-IPN, SIP-20151031
N. B.acknowledges financial support by Conacyt through the project 166581.
E. Rojas acknowledges partial support from PROMEP, UV-CA-320: \'Algebra, Geometr\'\i a
y Gravitaci\'on.
\end{acknowledgments}

\bibliography{biblio2_good_A}

\begin{thebibliography}{35}
\expandafter\ifx\csname natexlab\endcsname\relax\def\natexlab#1{#1}\fi
\expandafter\ifx\csname bibnamefont\endcsname\relax
  \def\bibnamefont#1{#1}\fi
\expandafter\ifx\csname bibfnamefont\endcsname\relax
  \def\bibfnamefont#1{#1}\fi
\expandafter\ifx\csname citenamefont\endcsname\relax
  \def\citenamefont#1{#1}\fi
\expandafter\ifx\csname url\endcsname\relax
  \def\url#1{\texttt{#1}}\fi
\expandafter\ifx\csname urlprefix\endcsname\relax\def\urlprefix{URL }\fi
\providecommand{\bibinfo}[2]{#2}
\providecommand{\eprint}[2][]{\url{#2}}

\bibitem[{\citenamefont{Amendola and Tsujikawa}(2010)}]{Amendo}
\bibinfo{author}{\bibfnamefont{L.}~\bibnamefont{Amendola}} \bibnamefont{and}
  \bibinfo{author}{\bibfnamefont{S.}~\bibnamefont{Tsujikawa}},
  \emph{\bibinfo{title}{Dark Energy:Theory and Observations}}
  (\bibinfo{publisher}{Cambridge Universirty Press}, \bibinfo{year}{2010}).

\bibitem[{\citenamefont{Arkani-Hamed et~al.}(1998)\citenamefont{Arkani-Hamed,
  Dimopoulos, and Dvali}}]{ArkaniHamed:1998rs}
\bibinfo{author}{\bibfnamefont{N.}~\bibnamefont{Arkani-Hamed}},
  \bibinfo{author}{\bibfnamefont{S.}~\bibnamefont{Dimopoulos}},
  \bibnamefont{and} \bibinfo{author}{\bibfnamefont{G.}~\bibnamefont{Dvali}},
  \bibinfo{journal}{Phys.Lett.} \textbf{\bibinfo{volume}{B429}},
  \bibinfo{pages}{263} (\bibinfo{year}{1998}), \eprint{hep-ph/9803315}.

\bibitem[{\citenamefont{Antoniadis et~al.}(1998)\citenamefont{Antoniadis,
  Arkani-Hamed, Dimopoulos, and Dvali}}]{Antoniadis:1998ig}
\bibinfo{author}{\bibfnamefont{I.}~\bibnamefont{Antoniadis}},
  \bibinfo{author}{\bibfnamefont{N.}~\bibnamefont{Arkani-Hamed}},
  \bibinfo{author}{\bibfnamefont{S.}~\bibnamefont{Dimopoulos}},
  \bibnamefont{and} \bibinfo{author}{\bibfnamefont{G.}~\bibnamefont{Dvali}},
  \bibinfo{journal}{Phys.Lett.} \textbf{\bibinfo{volume}{B436}},
  \bibinfo{pages}{257} (\bibinfo{year}{1998}), \eprint{hep-ph/9804398}.

\bibitem[{\citenamefont{Randall and
  Sundrum}(1999{\natexlab{a}})}]{Randall:1999ee}
\bibinfo{author}{\bibfnamefont{L.}~\bibnamefont{Randall}} \bibnamefont{and}
  \bibinfo{author}{\bibfnamefont{R.}~\bibnamefont{Sundrum}},
  \bibinfo{journal}{Phys.Rev.Lett.} \textbf{\bibinfo{volume}{83}},
  \bibinfo{pages}{3370} (\bibinfo{year}{1999}{\natexlab{a}}),
  \eprint{hep-ph/9905221}.

\bibitem[{\citenamefont{Randall and
  Sundrum}(1999{\natexlab{b}})}]{Randall:1999vf}
\bibinfo{author}{\bibfnamefont{L.}~\bibnamefont{Randall}} \bibnamefont{and}
  \bibinfo{author}{\bibfnamefont{R.}~\bibnamefont{Sundrum}},
  \bibinfo{journal}{Phys.Rev.Lett.} \textbf{\bibinfo{volume}{83}},
  \bibinfo{pages}{4690} (\bibinfo{year}{1999}{\natexlab{b}}),
  \eprint{hep-th/9906064}.

\bibitem[{\citenamefont{Dvali et~al.}(2000)\citenamefont{Dvali, Gabadadze, and
  Porrati}}]{Dvali:2000hr}
\bibinfo{author}{\bibfnamefont{G.}~\bibnamefont{Dvali}},
  \bibinfo{author}{\bibfnamefont{G.}~\bibnamefont{Gabadadze}},
  \bibnamefont{and} \bibinfo{author}{\bibfnamefont{M.}~\bibnamefont{Porrati}},
  \bibinfo{journal}{Phys.Lett.} \textbf{\bibinfo{volume}{B485}},
  \bibinfo{pages}{208} (\bibinfo{year}{2000}), \eprint{hep-th/0005016}.

\bibitem[{\citenamefont{Cordero et~al.}(2012)\citenamefont{Cordero, Cruz,
  Molgado, and Rojas}}]{Cordero2012}
\bibinfo{author}{\bibfnamefont{R.}~\bibnamefont{Cordero}},
  \bibinfo{author}{\bibfnamefont{M.}~\bibnamefont{Cruz}},
  \bibinfo{author}{\bibfnamefont{A.}~\bibnamefont{Molgado}}, \bibnamefont{and}
  \bibinfo{author}{\bibfnamefont{E.}~\bibnamefont{Rojas}},
  \bibinfo{journal}{Class.Quant.Grav.} \textbf{\bibinfo{volume}{29}},
  \bibinfo{pages}{175010} (\bibinfo{year}{2012}), \eprint{1109.2332}.

\bibitem[{\citenamefont{Regge and Teiltelboim}(1977)}]{Regge}
\bibinfo{author}{\bibfnamefont{T.}~\bibnamefont{Regge}} \bibnamefont{and}
  \bibinfo{author}{\bibfnamefont{C.}~\bibnamefont{Teiltelboim}}, in
  \emph{\bibinfo{booktitle}{Proceedings of the Marcel Grossman Meeting,
  Trieste, Italy ,1975,edited by R. Ruffini}}
  (\bibinfo{publisher}{North-Holland, Amsterdam}, \bibinfo{year}{1977}),
  p.~\bibinfo{pages}{77}.

\bibitem[{\citenamefont{Davidson and Karasik}(1998)}]{Davidson:1999ru}
\bibinfo{author}{\bibfnamefont{A.}~\bibnamefont{Davidson}} \bibnamefont{and}
  \bibinfo{author}{\bibfnamefont{D.}~\bibnamefont{Karasik}},
  \bibinfo{journal}{Mod.Phys.Lett.} \textbf{\bibinfo{volume}{A13}},
  \bibinfo{pages}{2187} (\bibinfo{year}{1998}), \eprint{gr-qc/9901004}.

\bibitem[{\citenamefont{Karasik and Davidson}(2003)}]{Karasik:2002du}
\bibinfo{author}{\bibfnamefont{D.}~\bibnamefont{Karasik}} \bibnamefont{and}
  \bibinfo{author}{\bibfnamefont{A.}~\bibnamefont{Davidson}},
  \bibinfo{journal}{Phys.Rev.} \textbf{\bibinfo{volume}{D67}},
  \bibinfo{pages}{064012} (\bibinfo{year}{2003}), \eprint{gr-qc/0207061}.

\bibitem[{\citenamefont{Gurwich and Davidson}(2009)}]{Gurwich:2009rj}
\bibinfo{author}{\bibfnamefont{I.}~\bibnamefont{Gurwich}} \bibnamefont{and}
  \bibinfo{author}{\bibfnamefont{A.}~\bibnamefont{Davidson}},
  \bibinfo{journal}{Phys.Rev.} \textbf{\bibinfo{volume}{D80}},
  \bibinfo{pages}{024039} (\bibinfo{year}{2009}), \eprint{0907.3565}.

\bibitem[{\citenamefont{Cruz and Rojas}(2013)}]{CQG-2013}
\bibinfo{author}{\bibfnamefont{M.}~\bibnamefont{Cruz}} \bibnamefont{and}
  \bibinfo{author}{\bibfnamefont{E.}~\bibnamefont{Rojas}},
  \bibinfo{journal}{Class. Quant. Grav.} \textbf{\bibinfo{volume}{30}},
  \bibinfo{pages}{115012} (\bibinfo{year}{2013}).

\bibitem[{\citenamefont{Svetina and Zekz}(1989)}]{Svetina1989}
\bibinfo{author}{\bibfnamefont{S.}~\bibnamefont{Svetina}} \bibnamefont{and}
  \bibinfo{author}{\bibfnamefont{B.}~\bibnamefont{Zekz}},
  \bibinfo{journal}{Eur. Biophys.} \textbf{\bibinfo{volume}{17}},
  \bibinfo{pages}{101} (\bibinfo{year}{1989}).

\bibitem[{\citenamefont{Onder and Tucker}(1988)}]{Onder:1988wm}
\bibinfo{author}{\bibfnamefont{M.}~\bibnamefont{Onder}} \bibnamefont{and}
  \bibinfo{author}{\bibfnamefont{R.}~\bibnamefont{Tucker}},
  \bibinfo{journal}{J.Phys.} \textbf{\bibinfo{volume}{A21}},
  \bibinfo{pages}{3423} (\bibinfo{year}{1988}).

\bibitem[{\citenamefont{Cordero et~al.}(2011)\citenamefont{Cordero, Molgado,
  and Rojas}}]{Cordero:2010vv}
\bibinfo{author}{\bibfnamefont{R.}~\bibnamefont{Cordero}},
  \bibinfo{author}{\bibfnamefont{A.}~\bibnamefont{Molgado}}, \bibnamefont{and}
  \bibinfo{author}{\bibfnamefont{E.}~\bibnamefont{Rojas}},
  \bibinfo{journal}{Class.Quant.Grav.} \textbf{\bibinfo{volume}{28}},
  \bibinfo{pages}{065010} (\bibinfo{year}{2011}), \eprint{1012.1379}.

\bibitem[{\citenamefont{de~Rham and Tolley}(2010)}]{deRham:2010}
\bibinfo{author}{\bibfnamefont{C.}~\bibnamefont{de~Rham}} \bibnamefont{and}
  \bibinfo{author}{\bibfnamefont{A.~J.} \bibnamefont{Tolley}},
  \bibinfo{journal}{JCAP} \textbf{\bibinfo{volume}{1005}}, \bibinfo{pages}{015}
  (\bibinfo{year}{2010}), \eprint{1003.5917}.

\bibitem[{\citenamefont{Goon et~al.}(2011)\citenamefont{Goon, Hinterbichler,
  and Trodden}}]{Trodden2011}
\bibinfo{author}{\bibfnamefont{G.}~\bibnamefont{Goon}},
  \bibinfo{author}{\bibfnamefont{K.}~\bibnamefont{Hinterbichler}},
  \bibnamefont{and} \bibinfo{author}{\bibfnamefont{M.}~\bibnamefont{Trodden}},
  \bibinfo{journal}{JCAP} \textbf{\bibinfo{volume}{017}}, \bibinfo{pages}{1107}
  (\bibinfo{year}{2011}).

\bibitem[{\citenamefont{Janet}(1926)}]{Janet:1926}
\bibinfo{author}{\bibfnamefont{M.}~\bibnamefont{Janet}},
  \bibinfo{journal}{Ann.Soc.Math.} \textbf{\bibinfo{volume}{5}},
  \bibinfo{pages}{38} (\bibinfo{year}{1926}).

\bibitem[{\citenamefont{Cartan}(1927)}]{Cartan:1927}
\bibinfo{author}{\bibfnamefont{E.}~\bibnamefont{Cartan}},
  \bibinfo{journal}{Ann.Soc.Math.} \textbf{\bibinfo{volume}{6}},
  \bibinfo{pages}{1} (\bibinfo{year}{1927}).

\bibitem[{\citenamefont{Clarke}(1970)}]{Clarke:1970}
\bibinfo{author}{\bibfnamefont{J.}~\bibnamefont{Clarke}},
  \bibinfo{journal}{Proc.R.Soc.London} \textbf{\bibinfo{volume}{A314}},
  \bibinfo{pages}{417} (\bibinfo{year}{1970}).

\bibitem[{\citenamefont{Rosen}(1965)}]{Rosen:1965}
\bibinfo{author}{\bibfnamefont{J.}~\bibnamefont{Rosen}}, \bibinfo{journal}{Rev.
  Mod. Phys.} \textbf{\bibinfo{volume}{37}}, \bibinfo{pages}{204}
  (\bibinfo{year}{1965}).

\bibitem[{\citenamefont{Cordero et~al.}(2014)\citenamefont{Cordero, Cruz,
  Molgado, and Rojas}}]{Cordero:2013pua}
\bibinfo{author}{\bibfnamefont{R.}~\bibnamefont{Cordero}},
  \bibinfo{author}{\bibfnamefont{M.}~\bibnamefont{Cruz}},
  \bibinfo{author}{\bibfnamefont{A.}~\bibnamefont{Molgado}}, \bibnamefont{and}
  \bibinfo{author}{\bibfnamefont{E.}~\bibnamefont{Rojas}},
  \bibinfo{journal}{Gen.Rel.Grav.} \textbf{\bibinfo{volume}{46}},
  \bibinfo{pages}{1761} (\bibinfo{year}{2014}), \eprint{1309.3031}.

\bibitem[{\citenamefont{Dunkley et~al.}(2005)\citenamefont{Dunkley, Bucher,
  Ferreira, Moodley, and Skordis}}]{Dunkley05}
\bibinfo{author}{\bibfnamefont{J.}~\bibnamefont{Dunkley}},
  \bibinfo{author}{\bibfnamefont{M.}~\bibnamefont{Bucher}},
  \bibinfo{author}{\bibfnamefont{P.~G.} \bibnamefont{Ferreira}},
  \bibinfo{author}{\bibfnamefont{K.}~\bibnamefont{Moodley}}, \bibnamefont{and}
  \bibinfo{author}{\bibfnamefont{C.}~\bibnamefont{Skordis}},
  \bibinfo{journal}{Mon. Not. Roy. Astron. Soc.}
  \textbf{\bibinfo{volume}{356}}, \bibinfo{pages}{925} (\bibinfo{year}{2005}),
  \eprint{astro-ph/0405462}.

\bibitem[{\citenamefont{Berg}(2004)}]{Berg}
\bibinfo{author}{\bibfnamefont{B.}~\bibnamefont{Berg}},
  \emph{\bibinfo{title}{Markov Chain Monte Carlo Simulations And Their
  Statistical Analysis: With Web-based Fortran Code}}
  (\bibinfo{publisher}{World Scientific Publishing Company, Incorporated},
  \bibinfo{year}{2004}), ISBN \bibinfo{isbn}{9789812389350}.

\bibitem[{\citenamefont{MacKay}(2003)}]{MacKay}
\bibinfo{author}{\bibfnamefont{D.~J.~C.} \bibnamefont{MacKay}},
  \emph{\bibinfo{title}{Information Theory, Inference and Learning Algorithms}}
  (\bibinfo{publisher}{Cambrdige University Press}, \bibinfo{year}{2003}), ISBN
  \bibinfo{isbn}{0521642981}.

\bibitem[{\citenamefont{Neal}(1993)}]{Neal}
\bibinfo{author}{\bibfnamefont{R.~M.} \bibnamefont{Neal}}, \bibinfo{type}{Tech.
  Rep.} \bibinfo{number}{CRG-TR-93-1}, \bibinfo{institution}{Dept. of Computer
  Science, University of Toronto} (\bibinfo{year}{1993}).

\bibitem[{\citenamefont{Suzuki et~al.}(2012)\citenamefont{Suzuki, Rubin,
  Lidman, Aldering, Amanullah et~al.}}]{Union21}
\bibinfo{author}{\bibfnamefont{N.}~\bibnamefont{Suzuki}},
  \bibinfo{author}{\bibfnamefont{D.}~\bibnamefont{Rubin}},
  \bibinfo{author}{\bibfnamefont{C.}~\bibnamefont{Lidman}},
  \bibinfo{author}{\bibfnamefont{G.}~\bibnamefont{Aldering}},
  \bibinfo{author}{\bibfnamefont{R.}~\bibnamefont{Amanullah}},
  \bibnamefont{et~al.}, \bibinfo{journal}{Astrophys.J.}
  \textbf{\bibinfo{volume}{746}}, \bibinfo{pages}{85} (\bibinfo{year}{2012}),
  \eprint{astro-ph/1105.3470}.

\bibitem[{\citenamefont{Moresco et~al.}(2012)\citenamefont{Moresco, Verde,
  Pozzetti, Jimenez, and Cimatti}}]{Jimenez12}
\bibinfo{author}{\bibfnamefont{M.}~\bibnamefont{Moresco}},
  \bibinfo{author}{\bibfnamefont{L.}~\bibnamefont{Verde}},
  \bibinfo{author}{\bibfnamefont{L.}~\bibnamefont{Pozzetti}},
  \bibinfo{author}{\bibfnamefont{R.}~\bibnamefont{Jimenez}}, \bibnamefont{and}
  \bibinfo{author}{\bibfnamefont{A.}~\bibnamefont{Cimatti}},
  \bibinfo{journal}{JCAP} \textbf{\bibinfo{volume}{1207}}, \bibinfo{pages}{053}
  (\bibinfo{year}{2012}), \eprint{astro-ph/1201.6658}.

\bibitem[{\citenamefont{{Padmanabhan} et~al.}(2012)\citenamefont{{Padmanabhan},
  {Xu}, {Eisenstein}, {Scalzo}, {Cuesta}, {Mehta}, and
  {Kazin}}}]{Padmanabhan2012MNRAS}
\bibinfo{author}{\bibfnamefont{N.}~\bibnamefont{{Padmanabhan}}},
  \bibinfo{author}{\bibfnamefont{X.}~\bibnamefont{{Xu}}},
  \bibinfo{author}{\bibfnamefont{D.~J.} \bibnamefont{{Eisenstein}}},
  \bibinfo{author}{\bibfnamefont{R.}~\bibnamefont{{Scalzo}}},
  \bibinfo{author}{\bibfnamefont{A.~J.} \bibnamefont{{Cuesta}}},
  \bibinfo{author}{\bibfnamefont{K.~T.} \bibnamefont{{Mehta}}},
  \bibnamefont{and} \bibinfo{author}{\bibfnamefont{E.}~\bibnamefont{{Kazin}}},
  \bibinfo{journal}{Mon.Not.Roy.Astron.Soc.} \textbf{\bibinfo{volume}{427}},
  \bibinfo{pages}{2132} (\bibinfo{year}{2012}), \eprint{1202.0090}.

\bibitem[{\citenamefont{{Anderson} et~al.}(2012)\citenamefont{{Anderson},
  {Aubourg}, {Bailey}, {Bizyaev}, {Blanton}, {Bolton}, {Brinkmann},
  {Brownstein}, {Burden}, {Cuesta} et~al.}}]{Anderson2012MNRAS}
\bibinfo{author}{\bibfnamefont{L.}~\bibnamefont{{Anderson}}},
  \bibinfo{author}{\bibfnamefont{E.}~\bibnamefont{{Aubourg}}},
  \bibinfo{author}{\bibfnamefont{S.}~\bibnamefont{{Bailey}}},
  \bibinfo{author}{\bibfnamefont{D.}~\bibnamefont{{Bizyaev}}},
  \bibinfo{author}{\bibfnamefont{M.}~\bibnamefont{{Blanton}}},
  \bibinfo{author}{\bibfnamefont{A.~S.} \bibnamefont{{Bolton}}},
  \bibinfo{author}{\bibfnamefont{J.}~\bibnamefont{{Brinkmann}}},
  \bibinfo{author}{\bibfnamefont{J.~R.} \bibnamefont{{Brownstein}}},
  \bibinfo{author}{\bibfnamefont{A.}~\bibnamefont{{Burden}}},
  \bibinfo{author}{\bibfnamefont{A.~J.} \bibnamefont{{Cuesta}}},
  \bibnamefont{et~al.}, \bibinfo{journal}{Mon.Not.Roy.Astron.Soc.}
  \textbf{\bibinfo{volume}{427}}, \bibinfo{pages}{3435} (\bibinfo{year}{2012}),
  \eprint{1203.6594}.

\bibitem[{\citenamefont{Blake et~al.}(2012)\citenamefont{Blake, Brough,
  Colless, Contreras, Couch et~al.}}]{Blake:2012pj}
\bibinfo{author}{\bibfnamefont{C.}~\bibnamefont{Blake}},
  \bibinfo{author}{\bibfnamefont{S.}~\bibnamefont{Brough}},
  \bibinfo{author}{\bibfnamefont{M.}~\bibnamefont{Colless}},
  \bibinfo{author}{\bibfnamefont{C.}~\bibnamefont{Contreras}},
  \bibinfo{author}{\bibfnamefont{W.}~\bibnamefont{Couch}},
  \bibnamefont{et~al.}, \bibinfo{journal}{Mon.Not.Roy.Astron.Soc.}
  \textbf{\bibinfo{volume}{425}}, \bibinfo{pages}{405} (\bibinfo{year}{2012}),
  \eprint{1204.3674}.

\bibitem[{\citenamefont{Beutler et~al.}(2011)\citenamefont{Beutler, Blake,
  Colless, Jones, Staveley-Smith et~al.}}]{Beutler:2011hx}
\bibinfo{author}{\bibfnamefont{F.}~\bibnamefont{Beutler}},
  \bibinfo{author}{\bibfnamefont{C.}~\bibnamefont{Blake}},
  \bibinfo{author}{\bibfnamefont{M.}~\bibnamefont{Colless}},
  \bibinfo{author}{\bibfnamefont{D.~H.} \bibnamefont{Jones}},
  \bibinfo{author}{\bibfnamefont{L.}~\bibnamefont{Staveley-Smith}},
  \bibnamefont{et~al.}, \bibinfo{journal}{Mon.Not.Roy.Astron.Soc.}
  \textbf{\bibinfo{volume}{416}}, \bibinfo{pages}{3017} (\bibinfo{year}{2011}),
  \eprint{1106.3366}.

\bibitem[{\citenamefont{{Hinshaw} et~al.}(2013)\citenamefont{{Hinshaw},
  {Larson}, {Komatsu}, {Spergel}, {Bennett}, {Dunkley}, {Nolta}, {Halpern},
  {Hill}, {Odegard} et~al.}}]{Hinshaw2013}
\bibinfo{author}{\bibfnamefont{G.}~\bibnamefont{{Hinshaw}}},
  \bibinfo{author}{\bibfnamefont{D.}~\bibnamefont{{Larson}}},
  \bibinfo{author}{\bibfnamefont{E.}~\bibnamefont{{Komatsu}}},
  \bibinfo{author}{\bibfnamefont{D.~N.} \bibnamefont{{Spergel}}},
  \bibinfo{author}{\bibfnamefont{C.~L.} \bibnamefont{{Bennett}}},
  \bibinfo{author}{\bibfnamefont{J.}~\bibnamefont{{Dunkley}}},
  \bibinfo{author}{\bibfnamefont{M.~R.} \bibnamefont{{Nolta}}},
  \bibinfo{author}{\bibfnamefont{M.}~\bibnamefont{{Halpern}}},
  \bibinfo{author}{\bibfnamefont{R.~S.} \bibnamefont{{Hill}}},
  \bibinfo{author}{\bibfnamefont{N.}~\bibnamefont{{Odegard}}},
  \bibnamefont{et~al.}, \bibinfo{journal}{ApJS} \textbf{\bibinfo{volume}{208}},
  \bibinfo{eid}{19} (\bibinfo{year}{2013}), \eprint{1212.5226}.

\bibitem[{\citenamefont{Wang and Wang}(2013)}]{Wang:2013mha}
\bibinfo{author}{\bibfnamefont{Y.}~\bibnamefont{Wang}} \bibnamefont{and}
  \bibinfo{author}{\bibfnamefont{S.}~\bibnamefont{Wang}},
  \bibinfo{journal}{Phys.Rev.} \textbf{\bibinfo{volume}{D88}},
  \bibinfo{pages}{043522} (\bibinfo{year}{2013}), \eprint{1304.4514}.

\bibitem[{\citenamefont{Ade et~al.}(2014)}]{PlanckXVI}
\bibinfo{author}{\bibfnamefont{P.}~\bibnamefont{Ade}} \bibnamefont{et~al.}
  (\bibinfo{collaboration}{Planck}), \bibinfo{journal}{Astron.Astrophys.}
  \textbf{\bibinfo{volume}{571}}, \bibinfo{pages}{A16} (\bibinfo{year}{2014}),
  \eprint{1303.5076}.

\end{thebibliography}
\bibliographystyle{apsrev}

\end{document}